\documentclass[12pt,a4paper,final]{iopart}

\usepackage{iopams}  
\usepackage{graphicx}
\usepackage{cite}
\usepackage[breaklinks=true,colorlinks=true,linkcolor=blue,urlcolor=blue,citecolor=blue]{hyperref}
\usepackage{textcomp}

\begin{document}

\title[]{Propagation-assisted generation of intense few-femtosecond high-harmonic pulses}

\author{B Major$^{1,2,*}$, M Kretschmar$^{3,*}$, O Ghafur$^3$, A Hoffmann$^3$, K Kov\'acs$^{4}$, K Varj\'u$^{1,2}$, B Senfftleben$^3$, J T\"ummler$^3$, I Will$^3$, T Nagy$^3$, D Rupp$^{3,5}$, M J J Vrakking$^3$, V Tosa$^4$, and B Sch\"utte$^3$}

\address{$^1$ELI-ALPS, ELI-HU Non-Profit Ltd, Dugonics t\'er 13, Szeged 6720, Hungary}
\address{$^2$Department of Optics and Quantum Electronics, University of Szeged, D\'om t\'er 9, Szeged 6720, Hungary}
\address{$^3$Max-Born-Institut, Max-Born-Str. 2A, 12489 Berlin, Germany}
\address{$^4$National Institute for Research and Development of Isotopic and Molecular Technologies, Donat str. 67-103, 400293 Cluj-Napoca, Romania}
\address{$^5$Current address: ETH Zurich, John-von-Neumann-Weg 9, 8093 Zürich, Switzerland}
\address{$^*$These authors contributed equally to this work.}
\ead{Balazs.Major@eli-alps.hu}
\ead{Bernd.Schuette@mbi-berlin.de}

\begin{abstract}
The ongoing development of intense high-harmonic generation (HHG) sources has recently enabled highly nonlinear ionization of atoms by the absorption of at least 10 extreme-ultraviolet (XUV) photons within a single atom [Senfftleben \textit{et al.}, arXiv1911.01375]. Here we investigate the role that reshaping of the fundamental, few-cycle, near-infrared (NIR) driving laser within the 30-cm-long HHG Xe medium plays in the generation of the intense HHG pulses. Using an incident NIR intensity that is higher than what is required for phase-matched HHG, signatures of reshaping are found by measuring the NIR blueshift and the fluorescence from the HHG medium along the propagation axis. These results are well reproduced by numerical calculations that show temporal compression of the NIR pulses in the HHG medium. The simulations predict that after refocusing an XUV beam waist radius of 320\,nm and a clean attosecond pulse train can be obtained in the focal plane, with an estimated XUV peak intensity of $9\times 10^{15}$\,W/cm$^2$. Our results show that XUV intensities that were previously only available at large-scale facilities can now be obtained using moderately powerful table-top light sources. 
\end{abstract}

\vspace{2pc}
\noindent{\it Keywords}: high-harmonic generation, extreme-ultraviolet pulses, nonlinear optics, pulse reshaping, fluorescence
%\submitto{\JPP}
%\maketitle

\section{Introduction}

During the past two decades, extreme-ultraviolet (XUV) sources based on high-harmonic generation (HHG) with an intensity high enough to drive nonlinear processes have provided novel and exciting opportunities for the investigation of ultrafast phenomena on the shortest timescales~\cite{midorikawa08}. For instance, measurements of intense HHG pulses have revealed the attosecond pulse train structure of the radiation produced by HHG~\cite{tzallas03}, and it has been demonstrated that these pulses can be sufficiently powerful for material modification purposes~\cite{mashiko04}. Other important applications are XUV-XUV pump-probe experiments that study dynamics in atoms and molecules without the need to use a near-infrared (NIR) laser field that may strongly perturb the dynamics of interest. In these experiments, a time resolution between 1\,fs~\cite{tzallas11} and 500\,as~\cite{takahashi13} has been achieved. Recently, it has furthermore become possible to perform single-shot coherent diffractive imaging of isolated nanotargets using intense HHG pulses~\cite{rupp17}, a type of experiment that was previously only possible at large-scale free-electron lasers (FELs)~\cite{bogan08}.

Although HHG sources are far more abundant than FELs, the number of intense HHG sources is still rather limited, see e.g.~\cite{tzallas03, mashiko04, kobayashi98, ravasio09, schutte14, manschwetus16, bergues18, nayak18, senfftleben19}. The development of new facilities like the Extreme Light Infrastructure (ELI) projects in Szeged~\cite{kuhn17} and Prague~\cite{hort19}, however, will make intense HHG pulses available to a large user community. Furthermore, we may expect additional intense HHG sources to become operational in the near future, see e.g. Ref.~\cite{wang18}. Because of this development and the exciting opportunities that are provided by intense HHG sources, it is important to thoroughly investigate their properties. An important goal is to develop schemes that allow an enhancement of the XUV intensity that can be achieved, and that at the same time provide attosecond time resolution in the experimental region.

Recently, we have suggested a novel scheme for enhancing the focused intensity of XUV pulses based on HHG~\cite{senfftleben19}. According to this scheme and contrary to the way that intense HHG sources have typically been optimized so far, the XUV intensity in the experimental region can be optimized by choosing a distance between the HHG source and the XUV focusing mirror that is significantly longer than the distance between the NIR focusing mirror and the HHG source. In this way, both the XUV focal spot size and the XUV beamline transmission can be optimized. We have used these pulses to ionize neutral Ar atoms and have observed ions with charge states up to Ar$^{5+}$, requiring the absorption of at least 10 XUV photons~\cite{senfftleben19}. 

Working with a relatively short NIR focal length and applying a conventional HHG scheme would mean that only a small fraction of the NIR pulse energy delivered by existing terawatt-class driving lasers (see e.g.~\cite{wu13, rudawski13, kuhn17, nayak18, kretschmar19}) could be used for HHG. The underlying reason is that phase-matched HHG requires a certain NIR intensity that is given by the driving laser wavelength and the generation medium. A way to increase the usable NIR pulse energy within a given focusing geometry is to exploit reshaping of the driving laser in the HHG medium. This has been studied both for the generation of XUV~\cite{tamaki99,tosa03,sun17,rivas18,major19} and soft X-ray pulses~\cite{schutte15,johnson18}. It has been demonstrated in a number of these studies that the HHG flux can be significantly increased by exploiting propagation effects (see e.g.~\cite{tamaki99,sun17,johnson18}).  

In this paper, we demonstrate how reshaping of the NIR driving pulses in a 30-cm-long gas cell filled with Xe can be used to assist the generation of intense XUV pulses based on HHG. The idea is to apply an NIR pulse at an intensity that significantly exceeds the optimum NIR intensity for phase-matched HHG in Xe. Propagation effects in the gas medium result in a reduction of the NIR intensity towards the end of the gas cell, where high harmonics are efficiently generated. Experimentally, we have identified propagation effects both by measuring fluorescence from the HHG cell and by measuring a blueshift in the NIR spectrum following propagation of the NIR pulse through the HHG medium. These experimental observations are well reproduced by numerical calculations. The simulations further predict a clean attosecond pulse train in the focal plane of the experiment as well as a very small XUV beam waist radius of 320\,nm, leading to an estimated XUV peak intensity of $9 \times 10^{15}$\,W/cm$^2$ in the experimental region of our HHG beamline~\cite{senfftleben19}.

\section{Experimental setup}

\begin{figure}[htb!]
 \centering
	\fbox{\includegraphics[width=\linewidth]{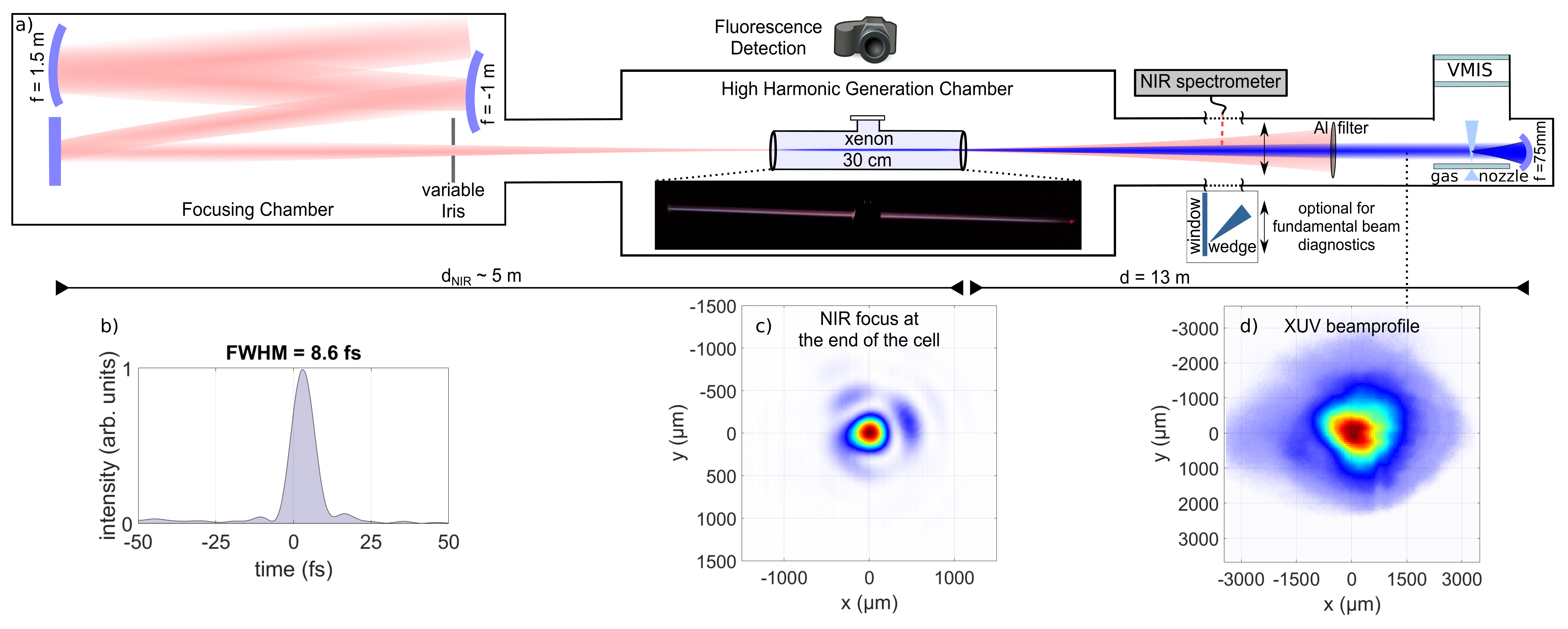}}
 \caption{\label{setup} (a) Experimental setup. NIR driving pulses with a duration of 8.6\,fs are focused using a telescope consisting of a concave ($f=1.5$\,m) and a convex mirror ($f=-1$\,m). The last NIR folding mirror is located about 5~m in front of the focal plane. A motorized iris behind the last NIR mirror is used to optimize the HHG yield. Harmonics are generated in a 30-cm-long glass cell that is filled with Xe and that ends 10\,cm in front of the NIR focal plane. The fluorescence yield from the HHG cell is recorded using a CCD camera with a microscope objective. The NIR and XUV pulses co-propagate over a distance of 12\,m, where a 100\,nm thick Al filter blocks the NIR light. The XUV beam profile can be recorded using an MCP / phosphor screen assembly. At a distance of 13\,m from the HHG cell, the XUV pulses are focused using a B$_4$C-coated spherical mirror with a focal length of 75\,mm. A velocity-map imaging spectrometer (VMIS) is used to record photoelectron spectra. The NIR spectrum can be recorded using an NIR spectrometer after coupling out the NIR pulses from the vacuum through a 3-mm-thick window. (b) Measured NIR pulse duration showing small pre- and postpulses, (c) NIR focus profile measured with a CCD camera, showing spatial sidelobes, and (d) HHG beam profile.}
\end{figure}

The experiments were performed at the Max-Born-Institut (MBI) in Berlin~\cite{senfftleben19}. An optical parametric chirped pulse amplification (OPCPA) system that was developed in-house was used to drive HHG. This system consists of a three-stage optical parametric amplifier, pumped by a home-built frequency-doubled Yb:YAG thin disk laser that operates at 100\,Hz repetition rate. Ultrashort seed pulses from a commercial Yb:KGW front-end are amplified to pulses with an energy of up to 42\,mJ and a duration of $<9$\,fs in a spectral window from 675\,nm to 1025\,nm~\cite{kretschmar19}. We note that the system was not operated at full power for the current experiments.

A schematic illustration of the HHG beamline is shown in Fig.~\ref{setup}(a). A telescope consisting of a concave ($f=1.5$\,m) and a convex mirror ($f=-1$\,m) was used to focus the NIR pulses. This configuration has the advantages that astigmatism can be compensated and that the distance from the last curved mirror to the focal plane can be easily adjusted. In the present work, the distance between the last NIR mirror and the NIR focal plane was about 5\,m. To optimize the HHG yield, the aperture of a motorized iris could be varied, and the optimal HHG yield was found for an iris diameter of 10\,mm (to be compared to a $1/e^2$ beam diameter of 12.5\,mm at this position). Under these conditions, the NIR pulse energy measured behind the iris was 11\,mJ. About 50\,$\%$ of the NIR pulse energy was contained in pre- and post-pulses (see Fig.~\ref{setup}(b)) or was contained in spatial side lobes that are outside of the main NIR focal spot (see Fig.~\ref{setup}(c)). We have estimated an NIR peak intensity of $5 \times 10^{14}$\,W/cm$^2$ in the focal plane. To optimize the HHG yield, we have used different gas cells with lengths of 10\,cm, 20\,cm, 30\,cm, 40\,cm and 50\,cm. The highest flux was found for the 30-cm-long gas cell that was statically filled with Xe using a backing pressure of about 7\,mbar. Gas dynamics simulations were performed, indicating that the pressure in the gas cell was uniform and had a value that is about one order of magnitude lower than the backing pressure. The entrance and exit apertures of the gas cell were covered by self-adhesive Al foils, through which small holes were drilled by the incident NIR driving laser. 

To block the NIR light co-propagating with the HHG beam, a 100\,nm thick Al filter was placed 12\,m downstream from the HHG cell. Another 0.5\,m downstream, i.e. at a distance of 12.5\,m from the HHG cell, the HHG beam profile (Fig.~\ref{setup}(d)) was measured using a microchannel plate (MCP) / phosphor screen assembly in combination with a charged-coupled device (CCD) camera. The full-width-at-half-maximum (FWHM) of the XUV beam at this position was about 1.7\,mm, corresponding to a divergence of only 0.14\,mrad. For the measurement of a photoelectron spectrum, the XUV pulses were focused into a velocity-map imaging spectrometer (VMIS)~\cite{eppink97} using a spherical B$_4$C-coated mirror with a focal length of 75\,mm. To this end, a pulsed and skimmed atomic jet was injected into the interaction zone from below.

To investigate NIR propagation effects in the HHG medium, we have used an HHG cell made out of glass. A CCD camera in combination with a microscope objective was used to record fluorescence emerging from the cell as a function of the NIR propagation distance within the cell. Scattered NIR light was attenuated using a bandpass filter in front of the camera. Correspondingly, the fluorescence was measured in a wavelength range extending from about 370 to 650\,nm. In addition, we have measured wavelength spectra of the transmitted NIR pulses behind the cell. To this end, the NIR pulses were coupled out of the vacuum using a 3\,mm thick glass window that was placed about 5\,m behind the HHG cell. A screen was placed in the NIR beam path, and the NIR spectrum was recorded using scattered light from the screen both for optimized HHG conditions and when the HHG cell was evacuated. We have found that the NIR spectrum recorded from the screen showed no signatures of spectral phase modulation when being compared to the NIR spectrum measured at the output of the laser system.

\section{Numerical model}

The numerical modeling was performed using an adapted version of the 3D non-adiabatic model described in detail in Refs.~\cite{tosa2005-1, tosa2005-2}. The calculations were carried out in three major steps: (1) propagation of the driving laser pulse in the interaction medium; (2) calculation of the single atom dipole-response for the interaction with the laser pulse; (3) construction and propagation of the harmonic field through linear optical elements to the far-field. Specific experimental arrangements can be incorporated into the model. 

(1) In order to accurately treat the propagation of the driving laser pulse, we solve the non-linear wave equation for the carrier wave. In this particular case we used as input the experimentally measured spectral intensity and phase at the beginning of the cell filled with Xe. The wave equation is solved in the frequency domain using a self-consistent iterative method and applying the paraxial approximation. The driving pulse propagates in a gas medium characterized by a refractive index which rapidly changes in time and space, including contributions from dispersion and absorption of neutral atoms, the optical Kerr effect, and plasma dispersion. The variations of the refractive index cause the reshaping of the pulse. This aspect is important and will be discussed in section~\ref{simulations}.

(2) The laser-atom interaction is treated in the framework of the strong-field approximation~\cite{lewenstein1994}. The non-linear dipole is obtained by solving the Lewenstein integral, which is the source of the macroscopic harmonic radiation. Ground-state depletion by ionization is taken into account using an empirical formula for the static field ionization rates of atoms given by the Ammosov–Delone–Krainov (ADK) theory~\cite{tong2005}.

(3) The macroscopic harmonic field builds up from the non-linear dipole and propagates in the same medium as the fundamental pulse. The wave equation describing the harmonics' propagation has the same structure as the wave equation for the driving pulse, and it is solved for every frequency component. Absorption and dispersion of the harmonics are taken into account. The analysis on the build-up of the harmonic field and on the coherence length of the generated radiation is supplemented by phase-matching calculations described in detail in Ref. \cite{vozzi2011}.

In order to compare the experimental pulse propagation characteristics with the simulations, the plasma fluorescence was calculated. The calculation method was described in Ref.~\cite{takahashi2003}, with the main assumption that the intensity of the fluorescence signal is proportional to the ion concentration. Assuming only single ionization of the atoms in the interaction region (confirmed for the present situation by simulations), the measured fluorescence becomes a measure of the electron density. 

The propagation of the harmonic field after it exits the generation medium was calculated using the ABCD-Bessel formalism~\cite{zalevsky1998} (or also called ABCD-Hankel transform~\cite{kovacs2017}). This approach allows one to analyze free propagation of light through any optical element that can be described in the framework of the ABCD formalism \cite{major2018}.

\section{Signatures of NIR propagation effects}

\begin{figure}[hb!]
 \centering
	\fbox{\includegraphics[width=\linewidth]{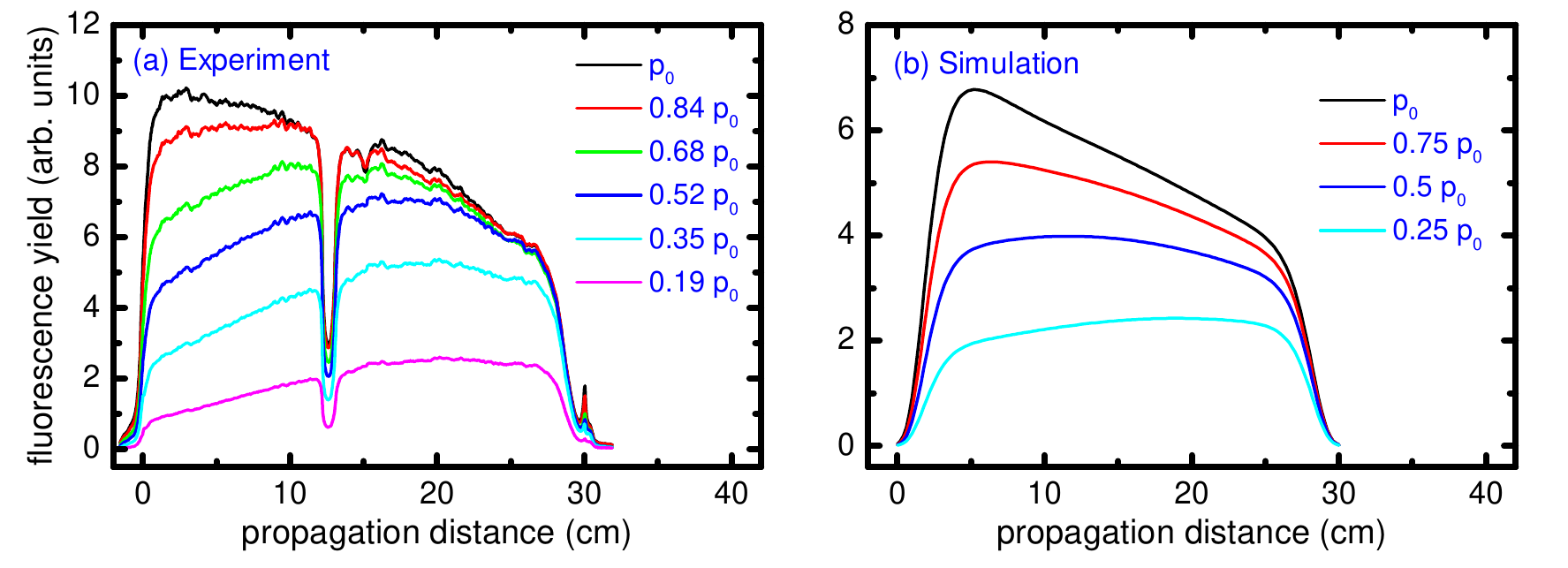}}
 \caption{\label{fluorescence} (a) Experimental fluorescence yield from the HHG cell measured as a function of the propagation distance of the NIR pulses within the cell, where the entrance of the cell is at 0\,cm and the exit of the cell is at 30\,~cm. Data are shown for different backing pressures, where $p_0$ corresponds to the pressure at which the highest HHG flux was observed. When increasing the pressure, the peak of the fluorescence yield shifts from close to the exit of the cell to close to the entrance of the cell. Note that the drop of the fluorescence signal in the region between 12\,cm and 13\,cm is due to the mounting of the cell and that the transmission of the glass cell decreases about 2~cm from the cell exit due to Al deposits on the interior wall of the cell caused by drilling of the Al foil that is attached to the cell exit by the NIR laser. The narrow peaks that are observed at the cell entrance and exit are due to scattered light at the Al foils. (b) Simulated fluorescence yields at various pressures, where $p_0=0.53$\,mbar.}
\end{figure}

NIR propagation effects in the HHG medium were studied by measuring the fluorescence yield in the glass cell as a function of the NIR propagation distance, see Fig.~\ref{fluorescence}(a). This experiment was performed at various backing pressures, and $p_0$ corresponds to the backing pressure where the highest HHG flux was observed. At this pressure, the fluorescence yield is peaked close to the entrance of the cell, whereas it is observed close to the exit of the cell at low pressures. This indicates reshaping of the NIR driving pulses within the HHG cell: At low pressures this effect is small, and the maximum fluorescence yield is observed close to the NIR focal plane, which was measured to be about 10~cm behind the exit of the cell when the cell was evacuated. At higher pressures, however, the generation of plasma within the cell leads to a situation where the NIR intensity is highest close to the entrance of the cell. The experimentally measured pressure dependence of the fluorescence yields is reproduced in the calculated propagation-dependent fluorescence yields presented in Fig.~\ref{fluorescence}(b). The underlying physical processes will be further discussed below. In the calculations, the optimum HHG yield was obtained at a pressure of $p_0=0.53$\,mbar. 

\begin{figure}[tb!]
 \centering
	\fbox{\includegraphics[width=\linewidth]{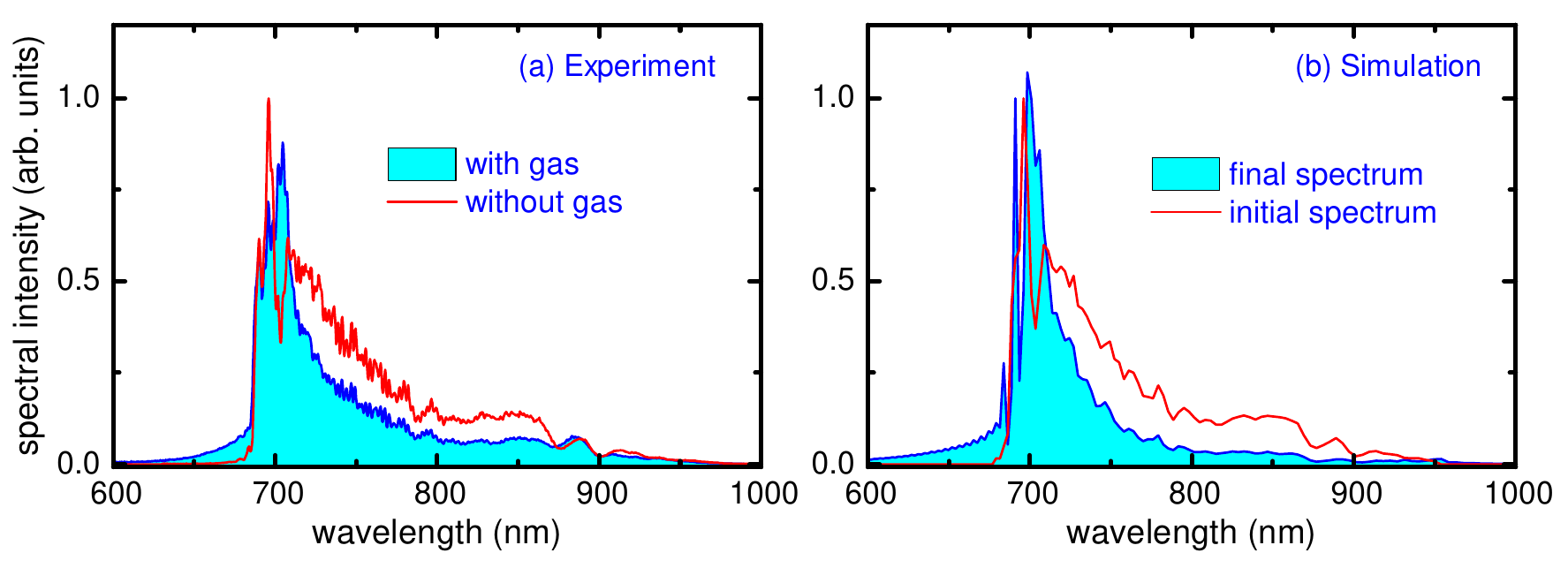}}
 \caption{\label{blueshift} NIR spectral blueshift. (a) NIR spectra behind the gas cell measured when the gas cell evacuated (red curve) and when the gas cell was filled with Xe at the optimal pressure for HHG (blue curve). A clear blueshift of the NIR spectrum is visible. Note that parts of the spectrum such as the peak around 885\,nm are not affected by the reshaping, indicating that these spectral components may be contained in a pre-pulse or in a spatial sidelobe in the focus (see Fig.\ref{setup}(b)+(c)). If so, these spectral components would not be exposed to a high plasma density and experience only modest change when propagating through the HHG cell. (b) NIR spectrum used as input in the simulation (red curve) and after propagation through the HHG medium (blue curve). The data were normalized to the integrated intensity and taking into account that about 20\,$\%$ of the radiation is absorbed.}
\end{figure}

We have further studied the NIR propagation effects by measuring NIR spectra behind the gas cell as depicted in Fig.~\ref{blueshift}(a). Here the red curve corresponds to the unperturbed spectrum that was measured when the cell was evacuated, and the blue curve corresponds to the spectrum that was obtained at the pressure where the maximum HHG flux was obtained. The observed blueshift induced by the gas shows that significant reshaping takes place. This is the result of a steep refractive index gradient in time due to plasma formation, which will be discussed in detail in Sec.~\ref{simulations}. The experimental results are qualitatively well reproduced by the simulations, as shown in Fig.~\ref{blueshift}(b). We note that the experimentally observed blueshift is somewhat smaller than the calculated blueshift. This might at least partially be attributed to the existence of the NIR pre- and postpulses (Fig.~\ref{setup}(b)), which are transmitted through the cell virtually unchanged, and the spatial sidelobes in the NIR focus (Fig.~\ref{setup}(c)), which might be exposed to a smaller plasma density.

\begin{figure}[htb!]
 \centering
	\fbox{\includegraphics[width=\linewidth]{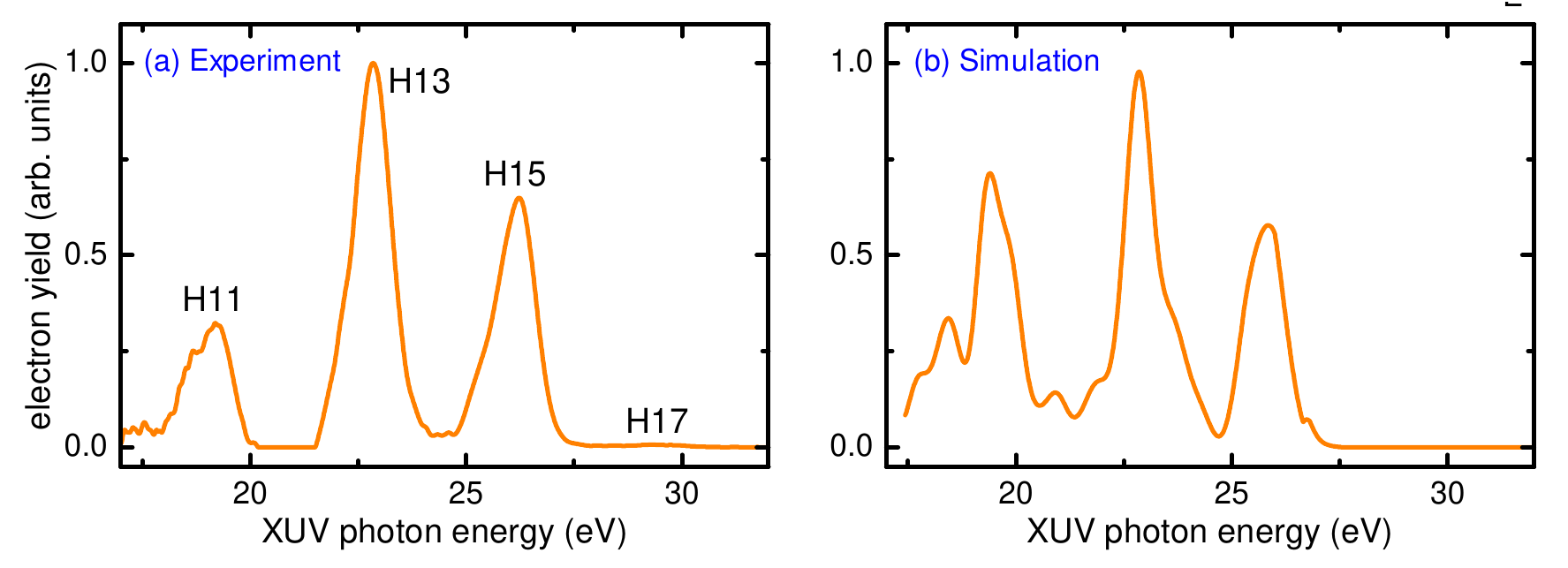}}
 \caption{\label{HHG-spectra}
 (a) Experimentally obtained photoelectron spectrum following the ionization of Ar atoms by the focused HHG pulse. The horizontal axis refers to the XUV photon energy scale, which has been obtained by adding the ionization potential of Ar ($\approx 15.8$\,eV) to the kinetic energies of the photoelectrons. (b) Calculated HHG spectrum taking into account the transmission of the Al filter, the reflectivity of the B$_4$C-coated mirror and the ionization cross sections of Ar.}
\end{figure}  

In Fig.~\ref{HHG-spectra}, we compare the photoelectron spectra obtained following the ionization of Ar atoms by the focused HHG pulses (see Fig.~\ref{HHG-spectra}(a)) and that we obtained numerically (see Fig.~\ref{HHG-spectra}(b)). In the latter case, the calculated HHG spectra were propagated to the interaction zone, taking into account the transmission of the Al filter \cite{henke}, the reflectivity of the XUV focusing mirror~\cite{larruquert98}, and the ionization cross sections of Ar atoms \cite{samson2002}. In both cases we find that the strongest contributions are from single-photon ionization by harmonics 11, 13 and 15. Furthermore, we observe a significant blueshift of the individual harmonics with respect to the photon energies that would be expected on the basis of the central frequency of the unperturbed NIR spectrum. This blueshift of the individual harmonics is a direct consequence of the blueshift experienced by the NIR pulse in the HHG cell and is consistent with the observed magnitude of this blueshift. The differences in the relative amplitudes of the individual harmonics can be attributed to a range of different effects, which include ageing of the Al filter, of the XUV focusing mirror and of the MCP detector used in the VMIS, which had a reduced sensitivity at lower photoelectron energies.

\section{Numerical results}
\label{simulations}
\subsection{Propagation-assisted harmonic generation}

\begin{figure}[htb!]
 \centering
	\fbox{\includegraphics[width=\linewidth]{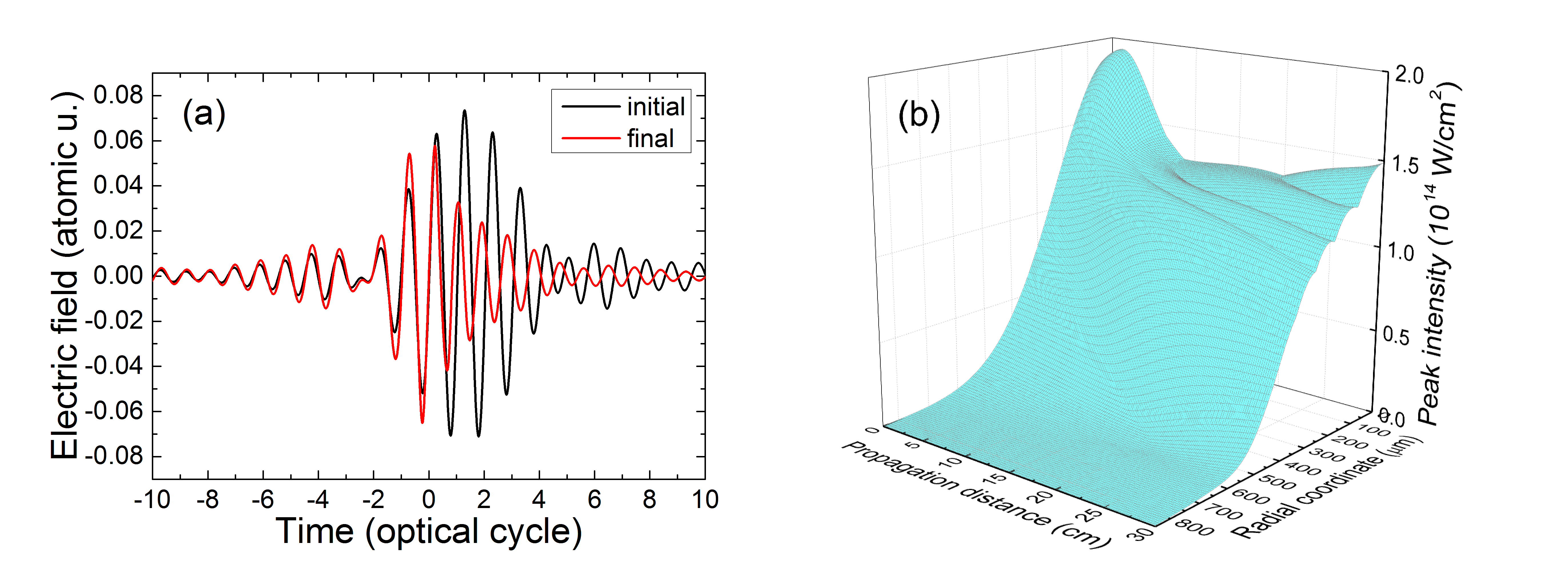}}
 \caption{\label{pulse_ampl} (a) Simulated on-axis driving field at the cell input and exit after 30\,cm propagation in a Xe medium at a pressure of 0.4\,torr. The input pulse shape was obtained from the experimentally measured spectral amplitude and phase. (b) Simulated spatial map of the driving pulse peak intensity in the gas medium.}
 \end{figure}

Fig.~\ref{pulse_ampl}(a) shows the simulated on-axis driving field at the cell input and at its exit, i.e. after propagation in the Xe medium at a pressure of 0.4\,torr and over a distance of 30\,cm.
One can see that reshaping takes place on the time-scale of one optical cycle via plasma-induced self-phase modulation (SPM)~\cite{bloembergen73} (note that this process is different from SPM based on the Kerr effect): The effective refractive index of the medium contains contributions from the dispersion of neutral atoms, the optical Kerr effect and plasma dispersion. As the leading edge of the driving pulse starts to ionize the medium, the plasma dispersion becomes the dominant effect which leads to rapid variation of the refractive index. As a consequence, the pulse experiences spectral reshaping (see Fig.~\ref{blueshift}) and temporal reshaping (red curve in Fig.~\ref{pulse_ampl}(a)). The spatial distribution of the field intensity is also affected, since the on-axis peak intensity decreases. However, the ionization level is much lower further away from the optical axis, and, as a consequence, also the reshaping is less pronounced. The result is the clamping of the NIR intensity over a radial range that exceeds the waist radius of the incident NIR laser pulse (Fig.~\ref{pulse_ampl}(b)), leading to favorable conditions for phase-matching over the entire beam profile increasing the HHG efficiency~\cite{major19}.

\begin{figure}[htb!]
 \centering
	\fbox{\includegraphics[width=\linewidth]{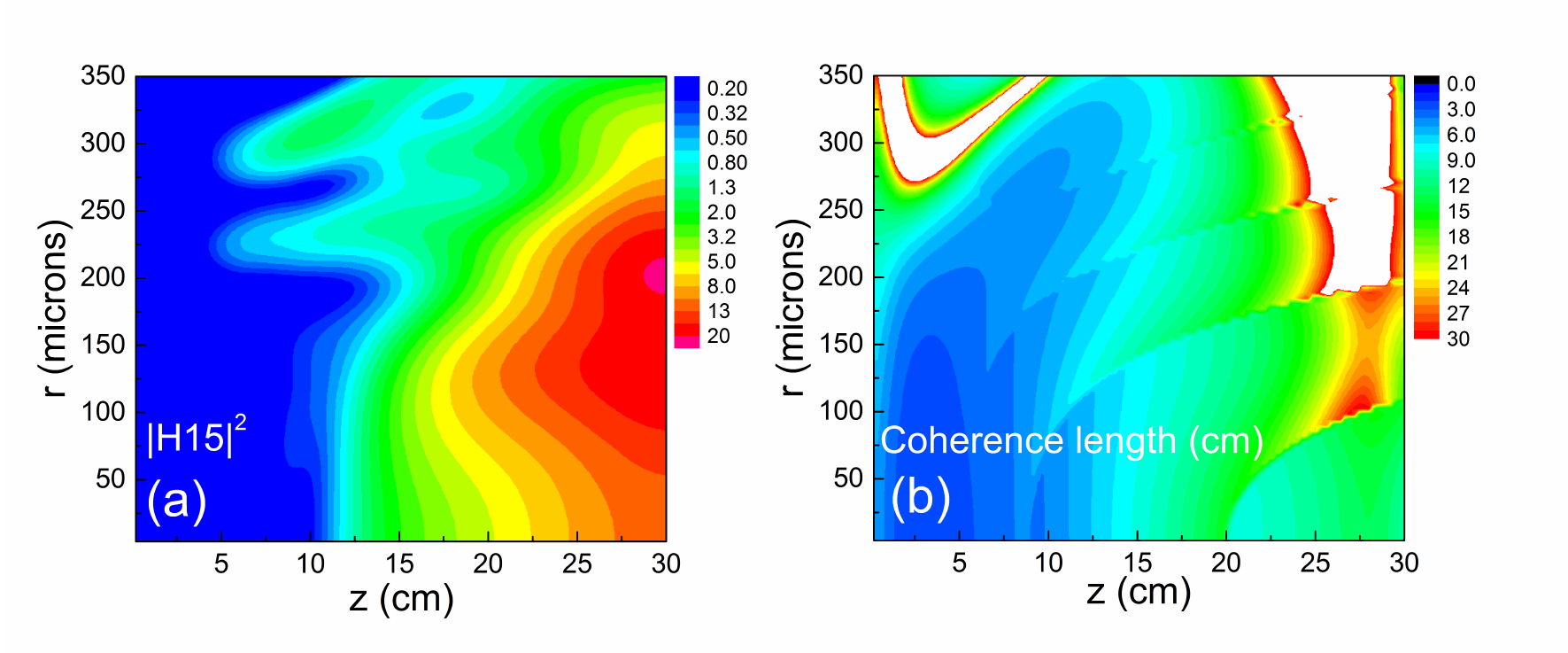}}
 \caption{\label{spnear} (a) Spatial build-up of H15 in the interaction volume, showing that H15 is dominantly generated in the second half of the cell and has an off-axis maximum in its intensity distribution. This plot was obtained by integration of the harmonic signal in a one harmonic-order-wide spectral window around the blueshifted H15 peak. We note that the HHG flux shows maximum at 30\,cm, in correspondence with the experimental observation. (b) Calculated coherence length map as a function of the propagation distance and the radial coordinate.}
 \end{figure}

Each of these modifications will affect the formation of the harmonics starting from the temporal / spectral characteristics of the atomic polarization and ending with the phase-matching conditions which will govern the pattern of the harmonic field build-up. For example, in this chain of effects the blueshift that the NIR laser undergoes in the cell (see Fig.~\ref{blueshift}) leads to a blueshift of the generated harmonics shown in Fig.~\ref{HHG-spectra}, i.e. an apparent increase of the harmonic order of the generated XUV radiation, which is calculated as the ratio between the XUV frequency and the carrier frequency of the incident NIR driving laser. Likewise, the spatial pattern of the driving field intensity (Fig.~\ref{pulse_ampl}(b)) will influence the dipole intensity map and phase-matching conditions, leading to the build-up of the harmonics as seen in Fig.~\ref{spnear}(a) for the intensity of the blueshifted harmonic H15: The maximum is observed at the end of the cell at an off-axis position. Fig.~\ref{spnear}(b) shows coherence length calculations, confirming that H15 is substantially better phase-matched in the second half of the cell than in the first half.

%(a) Simulated harmonic spectrum at different propagation distances within the cell. The horizontal axis is given in units of the harmonic orders of the unperturbed NIR photon energy, showing a clear blueshift of the individual harmonic orders.

\begin{figure}
 \centering
	\fbox{\includegraphics[width=0.5\textwidth]{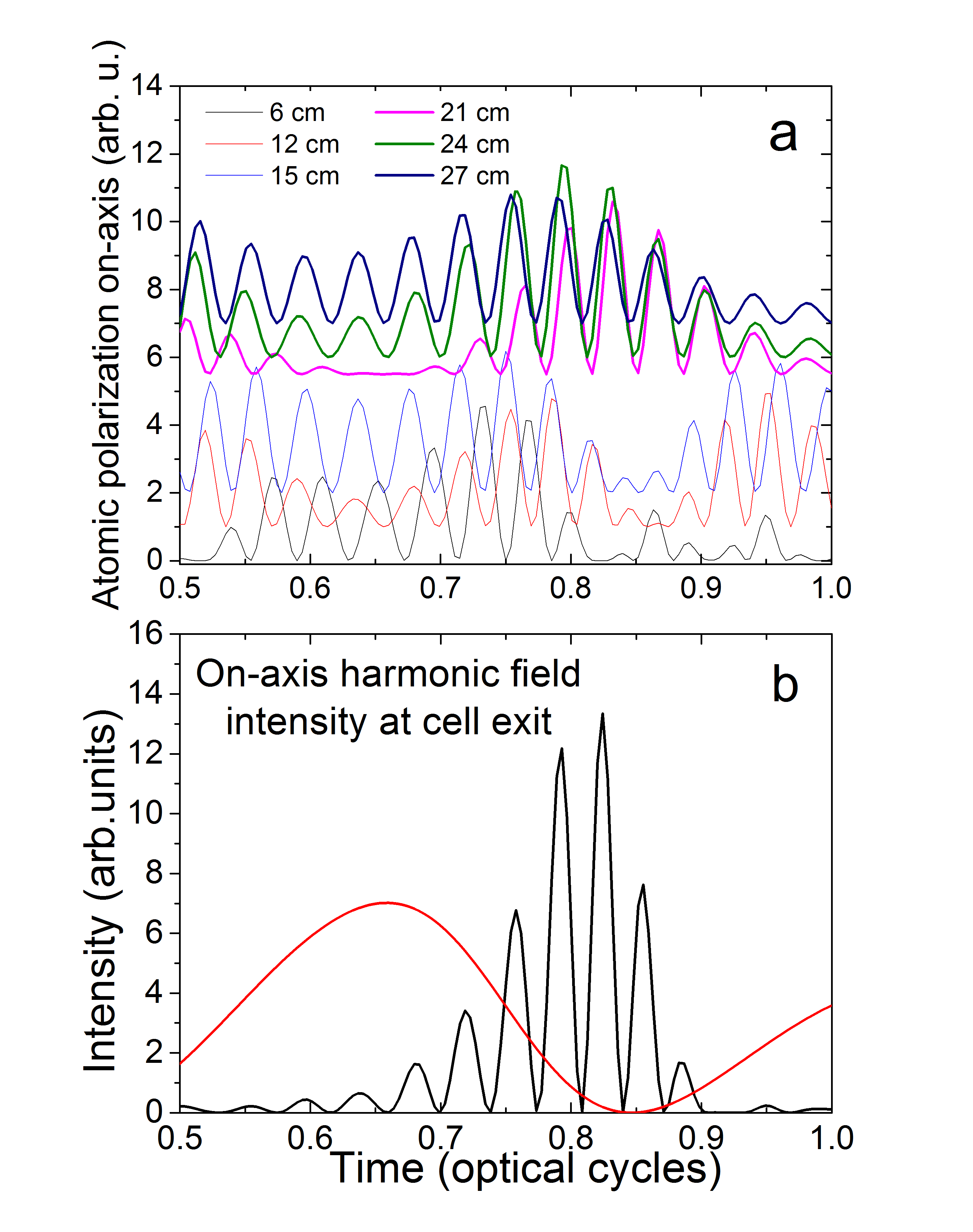}}
 \caption{\label{synchron} (a) On-axis atomic polarization at different propagation distances in the interaction medium. (b) The resulting harmonic field at the medium exit. The red curve shows the NIR driving field intensity, emphasizing that the build-up of one attosecond burst is plotted.}
 \end{figure}

The fact that phase-matching takes place in the second half of the medium is illustrated in Fig.~\ref{synchron}, which represents an alternative picture of the phase-matching mechanism. Fig.~\ref{synchron}(a) shows the on-axis, spectrally filtered (within a window spanning from H11 to H23) time-dependent atomic polarization for different positions within the HHG gas cell. We observe how one attosecond burst of the XUV pulse train  builds up due to the coherent addition of all atomic polarization contributions during the propagation of the NIR driving laser field in the gas cell. Fig.~\ref{synchron}(b) depicts in one half-cycle temporal window (after the intensity maximum of the incoming NIR pulse, i.e. $t=0$) the resulting harmonic field at the medium exit. In Fig.~\ref{synchron}(a) the spectrally filtered atomic polarization varies strongly during propagation in the first half of the cell, precluding efficient coherent addition. In contrast, in the second half of the cell, phase-matching occurs and the phase of the generated XUV radiation only weakly changes with propagation distance, leading to a rapid growth of the harmonic yield.

In conclusion, the simulation results confirm and support the experimental finding that the long interaction domain is beneficial for obtaining high-flux harmonics. Both temporal / spectral and spatial reshaping of the driving pulse during the propagation contribute to the build-up of the macroscopic harmonic field. 
 
\subsection{Focusing of the XUV beam}

\begin{figure}[htb!]
 \centering
    	\fbox{\includegraphics[width=0.95\linewidth]{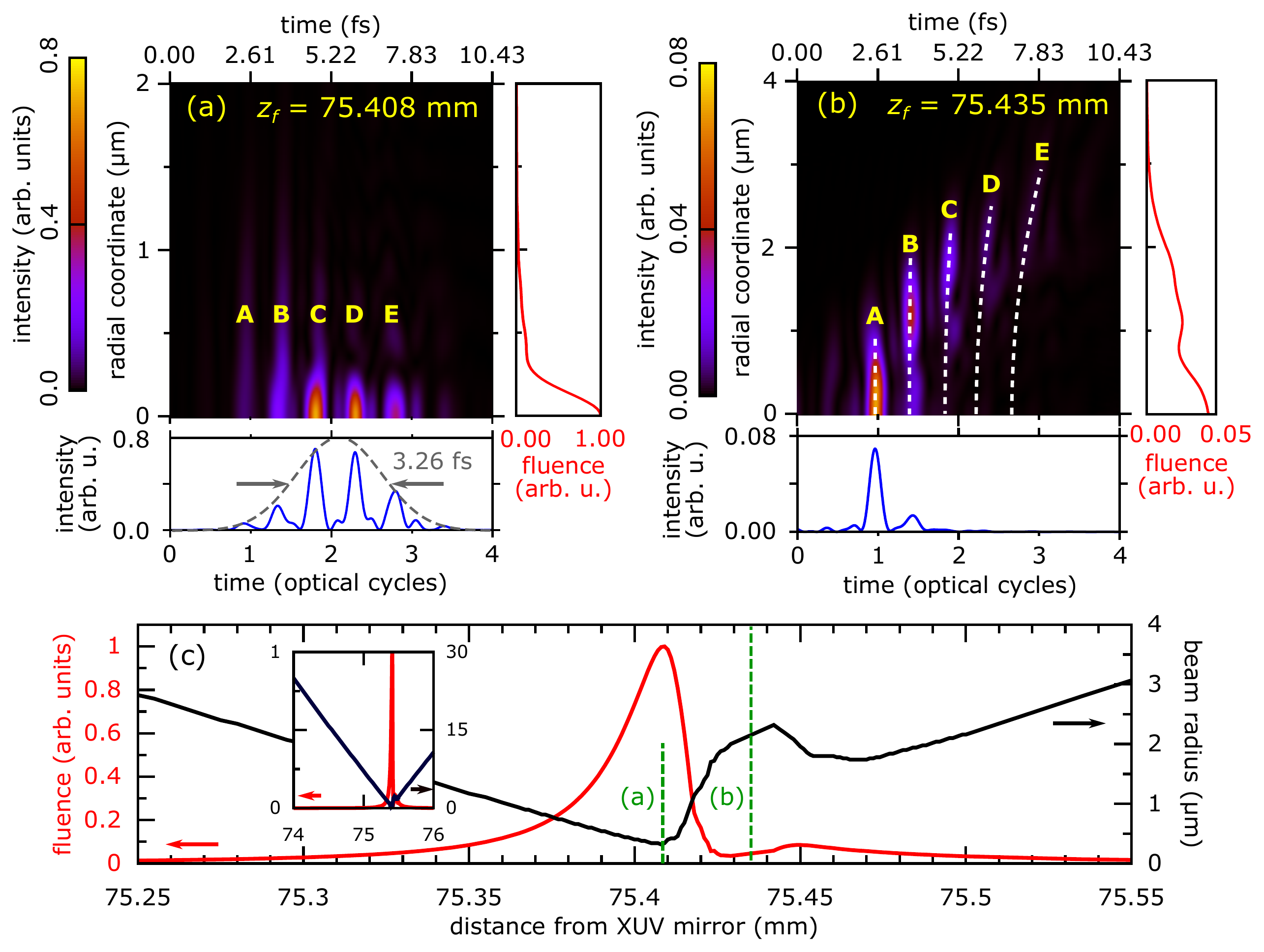}}
 \caption{\label{fig:XUVfoc}
 (a) The spatio-temporal structure of the high-harmonic radiation at the point of the highest peak fluence and smallest beam waist radius, which is $z_{f} = 75.408\,\mathrm{mm}$ away from the XUV focusing mirror. This corresponds to the image plane of the virtual HHG source. The subplot on the right shows the beam profile, which is the integrated signal along the time (horizontal) axis. The bottom subplot shows the temporal variation of the pulse train on axis, i.e. the lineout of the spatio-temporal map at $r = 0\,$\textmu m.
 (b) The same properties at $z_{f} = 75.435\,\mathrm{mm}$, which is the image plane of the cell end. The labels with capital letters indicate the five attosecond bursts of the pulse train originating from five consecutive half cycles of the laser field. The labels A, B, C, D and E in (a) and (b) refer to the individual attosecond bursts in the pulse train. The white dashed curves indicacte the wave fronts of the harmonic bursts.
 (c) The XUV peak fluence and beam radius (defined as $1/e^2$ radius) as a function of the distance from the focusing spherical mirror. The inset shows a zoom out with a wider range of propagation distances. The dashed vertical lines show the planes where (a) and (b) depict the spatio-temporal variation of the XUV field. The fluence and intensity values are in the same units throughout the figure.}
\end{figure}

We further carried out simulations to analyze the focusing properties of the XUV radiation in the VMIS chamber. These calculations took into account the transmission, dispersion and aperture of the Al filter, along with the reflectivity of the XUV spherical mirror (see experimental geometry in Fig.~\ref{setup}). The spatio-temporal structure in two representative planes behind the XUV spherical mirror are shown in Fig.~\ref{fig:XUVfoc}(a) and (b) (the planes are perpendicular to the propagation axis). Fig.~\ref{fig:XUVfoc}(a) shows the XUV radiation in the plane where the highest peak fluence is achieved, i.e. the XUV 'focus'. The position of the XUV focus is different from the nominal focus of the focusing mirror at $z_{f} = 75\,\mathrm{mm}$, because the generated XUV beam is divergent. In addition, this plane is not identical to the image plane of the end of the HHG cell shown in Fig.~\ref{fig:XUVfoc}(b). This is the result of a virtual source plane of the XUV beam, which we found to be $86\,\mathrm{cm}$ before the cell end using backpropagation. The virtual XUV beam waist radius at this position is 59\,$\mu$m, which is more than six times smaller than the XUV beam radius at the end of the cell. The existence of a virtual (or real) XUV focus, which is a result of the divergence (or convergence) of the harmonic beam after generation, has been identified earlier~\cite{frumker2012}, and has recently raised enhanced attention~\cite{wikmark2019, quintard2019}. The curved wave fronts of the XUV radiation exiting the generation medium can be seen in Fig.~\ref{fig:XUVfoc}(b) (indicated by white dashed curves), since this is a demagnified image of the cell end. It can further be seen that the attosecond bursts emerging from different half-cycles of the laser field have different divergences (see dashed curves in Fig.~\ref{fig:XUVfoc} and compare the radial positions of attosecond bursts with the same labels A, B, C, D and E in Fig.~\ref{fig:XUVfoc}(a) and (b)). The different curvatures of the attosecond pulses in the train is an indication of an ionization-induced attosecond lighthouse effect reported earlier~\cite{kovacs2017}. 
%The non-zero divergence of the radiation means that there is a virtual source of the XUV beam where the beam size would be minimal. 

By integrating the spatio-temporal signal along the time axis, the XUV beam profiles can be obtained. The peak fluence of the temporally integrated signal and the beam radius (defined as $1/e^2$ radius) can be seen in Fig.~\ref{fig:XUVfoc}(c) as a function of the propagation distance from the XUV spherical mirror. The inset of Fig.~\ref{fig:XUVfoc}(c) shows that the beam size changes linearly with the distance far from the XUV focus, like in the case of an ideal Gaussian beam. The divergence of the beam allows one to estimate the XUV beam waist radius. Assuming a central wavelength of $55.7\,\mathrm{nm}$ (corresponding to a photon energy of $22.2\,\mathrm{eV}$, which is the central photon energy of the radiation reflected from the spherical mirror, see Fig.~\ref{HHG-spectra}) and the divergence of $17.6\,\mathrm{mrad}$ (from the inset of Fig.~\ref{fig:XUVfoc}(a)), the XUV beam waist radius is $w_{\mathrm{XUV, foc}} = 1.01\,$\textmu m. 

%The corresponding Rayleigh length is $z_{\mathrm{XUV, foc}} = 57.62\,$\textmu m, which agrees well with the experimental estimation~\cite{senfftleben19}. 

When calculating the actual size of the focused XUV beam waist, however, we
obtain a more than three times smaller value of $w_{\mathrm{XUV, foc}} = 320$\,nm (see Fig.~\ref{fig:XUVfoc}). This value is significantly smaller than the value of $1.3$\,\textmu m that was estimated as an upper limit in Ref.~\cite{senfftleben19}. While it would not have been possible to resolve an XUV beam waist radius of 320\,nm due to the limited spatial resolution in~\cite{senfftleben19}, we note that the XUV mirror might further suffer from surface errors which were not taken into account in the calculations. The calculated XUV beam waist radius agrees well with the virtual XUV source radius of 59\,\textmu m at a distance of $86\,\mathrm{cm}$ before the cell end, when considering a demagnification factor of $D\approx d/f_{XUV}\approx 185$, where $d=13.86$\,m is the distance between the virtual XUV source size and the XUV focusing mirror, and $f_{XUV}=75$\,mm is the focal length of the XUV focusing mirror. This shows that the scaling behavior studied in Ref.~\cite{senfftleben19} may be applied also to non-Gaussian beams.

The difference between the focused beam sizes obtained based on the divergence of the beam and by calculating the spatio-temporal profile is a result of the divergence properties of the radiation from different half cycles. The far-field divergence is determined by the lowest-divergence half cycle (at 1 optical cycle in Fig.~\ref{fig:XUVfoc}(a) and (b)), while the other harmonics are generated in other half cycles with higher divergence, allowing for a smaller focused spot size, but they spread faster in the far field. The spurious beam-size increase close to the focus (between $75.41\,\mathrm{mm}$ and $75.47\,\mathrm{mm}$ in Fig.~\ref{fig:XUVfoc}(c)) is also attributed to these attosecond bursts with different divergences. Behind the focal plane, the higher-divergence half-cycles form an annular beam which leads to non-Gaussian (in a small region even to top-hat-like) beam profiles (see right subplot of Fig.~\ref{fig:XUVfoc}(b)).

Based on the simulated focused XUV beam waist radius ($w_{\mathrm{XUV, foc}} = 320$\,nm) and the pulse train duration ($\tau_{\mathrm{FWHM}} = 3.26\,\mathrm{fs}$, see pulse-train shape and fitted Gaussian envelope in bottom subplot of Fig.~\ref{fig:XUVfoc}(a)), the focused XUV peak intensity can be estimated to be $3\times10^{15}\,\mathrm{W/cm^{2}}$. Taking into account the attosecond structure of the pulse train, the peak intensity is $9\times10^{15}\,\mathrm{W/cm^{2}}$. This value takes into account that only $\sim 25$\% of the total beam energy (measured to be $70 \,\mathrm{nJ}$) is contained within the main Gaussian beam, while the rest of the energy is in the pedestals (see right subplot of the beam profile at focus in Fig.~\ref{fig:XUVfoc}(a)). 

\section{Conclusion and outlook}
In conclusion, we have shown that the XUV intensity that can be achieved using an HHG source can be substantially enhanced by exploiting the reshaping of the NIR driving laser in the HHG medium. In particular, three effects are important: (1) In a given NIR focusing geometry, the NIR pulse energy used for HHG can be significantly increased, since the optimal NIR intensity for phase-matched HHG is achieved after propagation in the HHG medium. Therefore, the radial range over which efficient HHG can take place exceeds significantly beyond the waist radius (in vacuum) of the incident NIR beam, and a higher HHG flux can be expected. (2) Propagation of the driving pulse results in phase-matching conditions leading to a comparably short attosecond pulse train. (3) Since the virtual HHG source radius is much smaller than the HHG beam radius at the end of the HHG cell, our simulations predict that a very small XUV beam waist radius of 320\,nm can be achieved after refocusing using a spherical mirror with a focal length of 75\,mm. While nanometer spot sizes from HHG sources have previously been achieved in the photon energy range around 90\,eV~\cite{ewald14, motoyama19}, this has to our knowledge not yet been achieved in the energy range around 15-30\,eV.

The estimated XUV intensity of $9 \times 10^{15}$~W/cm$^2$ opens the path to perform experiments, which have previously only been possible at free-electron lasers, using table-top light sources with few-femtosecond or even sub-femtosecond resolution. Examples include XUV-induced four-wave mixing~\cite{bencivenga15}, studies of superfluorescence~\cite{harries18} and the investigation of XUV-induced Stark shifts~\cite{ding19}. It might further become possible to study processes that were theoretically suggested, including transient impulsive stimulated Raman scattering~\cite{miyabe15} and the adiabatic passage to the continuum~\cite{saalmann18}. To make full use of the high XUV intensity, the short Rayleigh range of the focused XUV beam has to be taken into account. The XUV source is therefore ideally suited for experiments in solids and liquids. In gas-phase experiments, the use of very thin jets is promising, which might be achieved using nozzles with orifice diameters of several micrometers~\cite{braun97}. 

\section*{Acknowledgments}

The ELI-ALPS project (GINOP-2.3.6-15-2015-00001) is supported by the European Union and co-financed by the European Regional Development Fund. Funding by the Leibniz Grant No. SAW/2017/MBI4 is ackowledged. K. Kovacs and V. Tosa acknowledge support from a grant of the Romanian Ministry of Research and Innovation, CCCDI – UEFISCDI, project number PN-III-P1-1.2-PCCDI-2017-0010 / 74PCCDI ⁄ 2018, within PNCDI III. We acknowledge KIF\"U for awarding us access to HPC resources based in Hungary. We are grateful for the technical support by M. Krause, C. Reiter, W. Krüger and R. Peslin.

\section*{References}
%\bibliography{Bibliography}
\providecommand{\newblock}{}

\end{document}